**Disentangling magnetic hardening and molecular spin chain contributions to exchange bias in ferromagnet/molecule bilayers**

By *Samy Boukari\*, Hashim Jabbar, Filip. Schleicher, Manuel Gruber , Jacek Arabski, Victor Da Costa, Guy Schmerber, Prashanth Rengasamy, Bertrand Vileno, Wolfgang Weber, Martin Bowen\** and *Eric Beaurepaire\**

Dr. S. Boukari, Dr. H. Jabbar, Dr. F. Schleicher, Manuel Gruber[+], J. Arabski , Dr. V. Da Costa, Guy Schmerber, P. Rengasamy, Prof. W. Weber, Dr. M. Bowen and Dr. Eric Beaurepaire
Institut de Physique et Chimie des Matériaux de Strasbourg
Université de Strasbourg, CNRS UMR 7504
23 rue du Loess, BP 43, F-67034 Strasbourg Cedex 2 (France)
\* : e-mail: samy.boukari@ipcms.unistra.fr, martin.bowen@ipcms.unistra.fr , eric.beaurepaire@ipcms.unistra.fr
[+]Present address : Institut für Experimentelle und Angewandte Physik, Christian-Albrechts-Universität zu Kiel, 24098 Kiel, Germany
Dr. B. Vileno
Institut de Chimie de Strasbourg,
Université de Strasbourg, CNRS UMR7177
4 rue Blaise Pascal, F-67081 Strasbourg Cedex (France)
and
French EPR Federation of Research (REseau NAtional de Rpe interDisciplinaire (RENARD), Fédération IR-RPE CNRS 3443), France.

A paradigmic change in our understanding of the spin-polarized solid-state tunneling process (sp-SST) took place in the late 1990s-early 2000s, as the impact of the inorganic barrier's electronic structure (impact of *d* sites, structural ordering) on sp-SST was revealed against a backdrop of research using amorphous $Al_2O_3$ barriers. The giant jump in spintronic performance, with an effective spin polarization *P* that reaches 87% at room temperature (RT), reflects harnessing the conservation of electronic symmetry during SST[1], as well as the beneficial impact of oxygen vacancies according to recent research[2–4].

A similar paradigmic change is presently underway regarding organic tunnel barriers. Indeed, an initial understanding of sp-SST was established in the late 2000s across amorphous tunnel barriers such as those with $Alq_3$ molecules[5]. Here again, the barrier was first thought to merely constitute a means of spin transport between decoupled ferromagnetic (FM) electrodes. However, due to sizeable charge transfer, new electronic states with quite promising spintronic properties can occur at the interface between a ferromagnet and a molecule. Notably, a high spin polarization $P$ of the ferromagnet/molecule interface was detected using spin-polarized photoemission at RT on a wide variety of interface constituents[6,7], including amorphous carbon[8]. However, this high $P$ has thus far been evidenced in a solid-state device only at low temperature[9] (up to 99% at 2 K[10]), while similar measurements of $P$ at RT do not exceed 28%[5].

Unlocking the RT spintronic potential of the FM/molecule interface requires a better understanding of the magnetic exchange effects within FM/molecular films that involve this interface, which is also called an organic/molecular spinterface. Thus, far, two such effects have been identified: 1) a *magnetic hardening* of the topmost FM monolayer forming the organic spinterface compared to the underlying FM thin film, and 2) the impact on the organic spinterface's magnetic properties of *molecular spin chains* away from the interface and into the organic layer. Indeed, the adsorption of a molecule onto a FM surface promotes an interface with distinct magnetic properties relative to those of the FM substrate, both on the molecule[6,11,12] and on the topmost FM layer[13]. 1) On the metal side of the interface, this interface may have an increased anisotropy and a magnetization parallel or antiparallel to that of the FM substrate. This effect, called magnetic hardening[14,15], can in turn lead to magnetoresistance at RT[13]. However, the generality of the magnetic hardening effect remains unclear as it was demonstrated only for FM=Co and a radical molecule (zinc methyl phenalenyl). It is speculated that this magnetic hardening effect, depending on the strength of the interaction with the FM substrate, may lead to other macroscopic phenomena. In one such phenomenon, called exchange bias (EB), the center of the FM's hysteresis loop is shifted by a magnetic field $H_{shift}$ away from H=0. In a scenario involving only magnetic hardening, $H_{shift}$ should not depend on the thickness of the molecular film deposited atop the FM layer.

Since the magnetization of the first molecular monolayer (ML) can be fixed at RT through magnetic interactions with the FM substrate[6,11,12,16,17], this can in particular stabilize a molecule's intrinsic local magnetic moment. Furthermore, due to structural ordering, a magnetic order can occur between these local magnetic moments so as to form spin chains within a molecular film [18,19]. 2) Combining these two concepts, the spinterface's magnetism can stabilize[16] at RT an antiferromagnetic (AFM) ordering between the paramagnetic centers M of a metal phthalocyanine (MPc, schematized in Fig. 1(c)) film, thereby stabilizing the magnetic axis of a paramagnetic molecular spin chain. Conversely, at low temperature, the magnetization loop of the Co/MnPc bilayer exhibits the phenomenological feature of EB[20]: an anomalously large coercive field associated with a shift $H_{shift}$ in the loop center away from H=0[16]. This association is natural since EB, first discovered using inorganic materials, involves FM/AFM bilayers. In this scenario, we expect an increase of EB with the molecular film thickness up to several nm, which is typical for inorganic materials[21]. The EB variation would here be correlated with the length of the spin chains. As a note, Serri et al. found a lower limit of 5 nm for the spin chains length in CoPc films[19]

Thus, although this observation of EB is consistent with an explanation in terms of an effective FM/AFM magnetic ordering[16] of the Co/MnPc bilayer, it could also be explained in terms of magnetic hardening effects[13], which conceptually does not require an AFM ordering of molecular spin chains within the organic layer. Resolving this controversy is of prime importance to understand the origin of exotic magnetotransport results across Pc magnetic tunnel junctions[10,22] with Co electrodes: concurrent tunneling magnetoresistance / tunneling anisotropic magnetoresistance (TMR/TAMR); very high (up to 2T) bias-dependent coercive fields; concurrent unidirectional/uniaxial magnetotransport; and spectroscopic features that depend on the macroscopic MTJ magnetization state and that were tentatively attributed to spin-flip spectroscopy[23] across these solid-state devices. If spinterface-stabilized molecular spin chains[16], rather than only magnetic hardening[13], underscore these spin-flip events within organic SST, then the *magnetic ordering* within the organic tunnel barrier would imply that it is *structurally* ordered. In that case, these first magnetotransport results across MPc tunnel barriers[10,22] would mark the onset of SST across *ordered* organic barriers as a key milestone in this paradigmic shift away from *amorphous* organic SST, with promising prospects for quantum physics.

To address this pivotal open question within progress on organic SST toward ordered tunnel barriers, we deploy Superconducting Quantum Interference Device (SQUID) and ferromagnetic resonance (FMR) experiments on FM/MPc (FM=Co, $Ni_{81}Fe_{19}$ i.e. Py; and M=Mn, Fe, Co, Zn) bilayers. Changing the M site so as to tune the MPc molecule's intrinsic magnetic moment to be zero (ZnPc) or non-zero (MnPc, FePc, CoPc)[19,24–26], while maintaining a similar adsorption geometry of this planar molecular family onto the FM surface, allows us to selectively suppress spin chains within the MPc film. We can thus elegantly address several questions. 1) *Is EB arising here from spinterface-stabilized molecular spin chains or from magnetic hardening[13]?* We find that these two effects are distinct and additive. 2) Considering the limited number of FM/molecule pairs that reportedly generate these effects*, can one observe these magnetic exchange effects using other FM and molecular candidates?* We reproduce the EB effect within Co/ZnPc, Co/FePc and Co/CoPc bilayers in addition to Co/MnPc. The data on Co/ZnPc are interpreted in terms of the magnetic hardening effect. We find similar effects upon replacing Co with Py. 3) *Is there a correlation between $H_{shift}$ and a) the presence of a magnetic moment on the MPc molecule's M site and b) the thickness of the FM and MPc layers?* We find that $H_{shift}$ a) trends with the amplitude of the molecule's local magnetic moment $M_s$ and b) increases with the thickness of the CoPc up to 10 nm. This confirms the impact of molecular spin chains on the EB at organic spinterfaces[16] (see point 1), and underscores the interfacial impact of the effect as regards the FM layer. 4) *How does the strength of the EB generated using molecules compare with that using inorganic materials?* We find that this strength can be increased by a factor of up to 2-20 when using molecular rather than inorganic materials (see supporting information). 5) *Can*

*experiments unambiguously confirm the change in anisotropy[13] that was put forward to explain the magnetic hardening effect ?* We experimentally confirm this prediction using FMR. 6) *Can reports[10,22] of exotic magnetotransport effects across MPc films be ascribed to the impact of molecular spin chains?* We infer from our magnetometry analsysis that molecular spin chains play an important role in promoting the EB effect within the MnPc and CoPc films that were used as tunnel barriers in these magnetotransport experiments.[10,22] This lends strong credence to the impact of magnetic ordering, and therefore structural ordering, on these exotic magnetotransport results, which thus constitute the onset of tunnelling across a structurally ordered organic tunnel barrier.

MPc molecules were deposited on Si//SiO$_2$(500 nm)/Co(10 nm) or on Si//SiO$_2$(500 nm)/Cr(15 nm)/Py(10 nm) and protected with a 10 nm gold layer. All materials were deposited by sublimation in ultra-high vacuum (UHV), except Py and Cr which were deposited by sputtering while remaining in UHV. Structural characterization is reported in the Supporting Information (SI).

We present in Figure 2 SQUID magnetometry results on Co/MPc (M=Mn, Co, Fe, Zn) bilayers. Magnetization loops at 2 K after H=+3 T field cooling (FC: panel (a)) reveal, for all Co/MPc bilayers considered, that the hysteresis loop center is shifted away from H=0 by a negative H$_{shift}$ after FC. We present its temperature dependence in panel (b). For all MPc, we find that H$_{shift}$ decreases with increasing T, and vanishes at T~100 K. Since we observe a non-zero H$_{shift}$ for ZnPc, a molecule with neither radical nor 3$d$ magnetism, i.e. with zero nominal magnetic moment, this implies that spin chains within the molecular film are not required to observe the magnetic pinning of the FM substrate. This could, however, reflect the magnetic hardening effect as reported by Raman et al.[13], which in turn can pin the magnetization of the underlying FM substrate.

Furthermore, we observe that H$_{shift}$ increases in amplitude upon increasing the magnetic moment on the molecule's central site M from zero for Zn, to 1 µ$_B$ for Co, to 2 µ$_B$ for Fe, to 3

$\mu_B$ for Mn [19,24–26] (see Fig. 2c): it is an order of magnitude stronger when Zn is replaced by Mn. Does this increase in $H_{shift}$ originate from the spin chains or from a variation in magnetic hardening when changing the molecules ? To distinguish between the two scenarios, we examine the impact on $H_{shift}$ of varying the thickness of the FM and CoPc layers. Referring to Fig. 3a, we find that $H_{shift}$ decreases as $1/t_{Co}$ as the thickness of the Co layer $t_{Co}$ is increased. This confirms the interfacial nature of the EB from the standpoint of the FM layer's magnetism.

Referring to Fig. 3b, we find that $H_{shift}$ increases upon increasing the CoPc thickness up to ~10 nm, and then decreases slightly up to ~20 nm. Since the Co sites within CoPc exhibit AFM correlations[19], this shows that AFM spin chains within the MPc film can enhance $H_{shift}$. Since MPc films with M=Co, Fe, Mn can exhibit AFM spin chains[19], and $H_{shift}$ trends with the molecule's local magnetic moment (i.e. is always larger than for ZnPc with nominally zero local magnetic moment), we infer that EB using spinterface-stabilized molecular spin chains[16] is a distinct and additive effect to that of magnetic hardening[13]. Note how this combination of increase and decrease in $H_{shift}$ with increasing CoPc thickness is typical of an EB effect involving an inorganic AFM material with a varying grain size[21]. The enduring variation of $H_{shift}$ for CoPc thicknesses up to 20nm thus implies that MSC with an effective length of at least 20nm are involved in the EB effect.

Our results thus expand the list of molecules that generate EB from two[13,16] to five. To test whether the FM metal Co is required to observe the magnetic exchange effect, we replaced it with Py, and measured first SQUID magnetization loops after FC down to a temperature T. We present in Fig. 4a the resulting temperature dependence of $H_{shift}$ for Cr(15nm)/Py(10nm)/MPc(10nm)/Au(10nm) with M=Fe, Zn alongside that of a Cr(15nm)/Py(10nm)/Cr(10nm)/Au(10nm) reference. While the reference sample generates a non-zero $H_{shift}$~15 Oe as expected since Cr is AFM[27], replacing the top Cr layer with

ZnPc(FePc) increases $H_{shift}$ by a factor of two(six). We thus also witness the distinct, additive nature of the magnetic hardening and EB effects when FM=Py.

To confirm the prediction[13,14] that a strong anisotropy underscores the magnetic hardening effect, we performed FMR measurements on gold-capped Py/FePc bilayers. Since Py films exhibit a FMR resonance field of around 1000 Oe for X-band excitation that is significantly larger than the effective coercive field of the bilayer at all temperatures, the FMR technique is well adapted to study the two magnetic exchange effects[28] --- magnetic hardening and spinterface-stabilized molecular spin chains --- considered here. We first compare in Fig. 4a the amplitude and temperature dependence of $H_{shift}$ extracted for Py/FePc bilayers from SQUID and FMR measurements. We find that they are very similar, as expected since FMR is a perturbation technique. We now extract the uniaxial anisotropy field $H_a$ from FMR measurements performed with a magnetic field applied first in the film plane along the direction of field cooling, and then along the sample normal (see SI for details). We compare in Fig. 4b the temperature dependence of $H_a$ for a Cr/Py/FePc stack against that of a Cr/Py/Cr reference stack. As expected, the weak EB contribution from the Cr layers in the reference stack (see previous discussion) generates a low, positive $H_a$ at all temperatures. This is indicative of an out-of-plane uniaxial anisotropy. We find that $H_a$ decreases nearly monotonously with decreasing temperature from 300 K to 2 K. Turning now to the Cr/Py/FePc stack, we find the same sign and amplitude of $H_a$ at 300 K. However, $H_a$ decreases with a rate that increases around the 100 K blocking temperature observed for the Py/FePc bilayer (see Fig. 4(a)), and reaches a large negative value of $H_a \sim -4000$ Oe. Our experiments thus explicitly link the magnetic exchange effect to the presence of a large uniaxial anisotropy, in agreement with predictions[13,14]. The exotic magnetotransport results across Pc magnetic tunnel junctions[10,22] are thus due to a conjunction of an increase in the interfacial magnetic anisotropy upon molecular adsorption, i.e. due to the magnetic hardening effect, and of AFM spin chains in the organic tunnel barrier. This confirms that these results represent the onset of sp-SST across a structurally ordered organic tunnel barrier.

To conclude, we've performed SQUID and FMR magnetometry experiments to clarify the relationship between two reported magnetic exchange effects within ferromagnetic metal/molecule bilayers: the magnetic hardening effect reported using a custom-tailored molecule[13], and spinterface-stabilized molecular spin chains[16], both of which can affect the magnetization reversal of the underlying FM thin film. To distinguish between the two effects, we tuned the magnetic moment of the central site of the metal phthalocyanine molecular family to selectively enhance or suppress the formation of spin chains within the molecular film. We find that both effects are distinct, and additive. In the process, we 1) extended the list of FM/molecule candidate pairs that are known to generate magnetic exchange effects, 2) experimentally confirmed the predicted[13] increase in anisotropy upon molecular adsorption; and 3) showed that using molecules with a local magnetic moment enhances magnetic exchange due to the impact on the magnetism of the organic spinterface of molecular spin chains within the organic layer away from the interface.

Since the involvement of molecular spin chains implies structural ordering, our work explicitly ascribes the exotic magnetotransport reported in 2015 and 2016[10,22] across MPc tunnel barriers as the onset in organic solid-state tunnelling from amorphous to ordered organic tunnel barriers. This not only constitutes an echo to the milestone from amorphous to ordered inorganic tunnelling spintronics starting in 2001[29], but paves the way for solid-state devices studies that exploit the quantum physical properties of spin chains. Here, the recent proposal[15] and experimental demonstration[30] that the organic spinterface can constitute an active component toward multifunctional electronics may be used to electrically alter the molecular spin chain's ground/excited state, so as to craft spin-polarized transport and thus promote novel device functionalities.

**Acknowledgements**

The authors thank M. Bailleul, M. Hehn and S. Heutz for discussions as well as A. Derory and Ch. Kieber for technical assistance. We acknowledge the French EPR Federation of Research (REseau NAtional de Rpe interDisciplinaire, RENARD, Fedération IR-RPE CNRS 3443), funding from the Institut Carnot MICA's 'Spinterface' grant, from the Agence Nationale de la Recherche ANR-09-JCJC-0137 and ANR-11-LABX-0058 NIE (Labex NIE), from the International Center for Frontier Research in Chemistry and from the Franco-German University. H. J. acknowledges a grant from the Iraqi Ministry of Higher Education and Campus France.

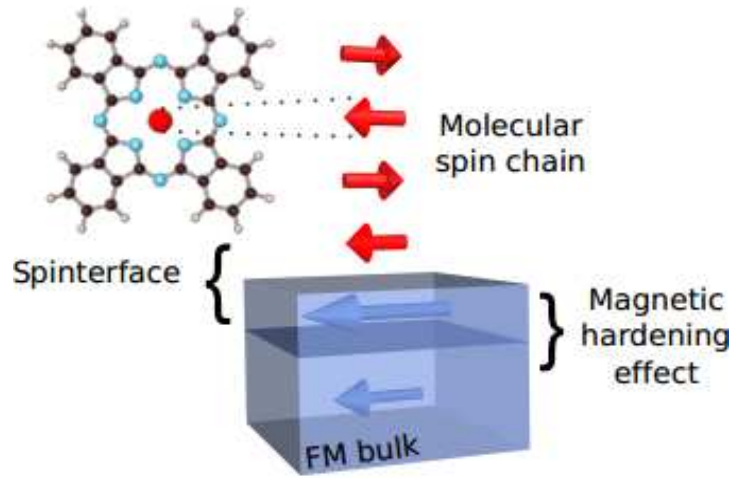

**Figure 1**. **Magnetic exchange effects within ferromagnet/molecule bilayers.** Spin-polarized charge transfer alters the properties of both molecule and ferromagnet upon adsorption. The magnetism of the resulting interface, also called an organic spinterface, differs from that of its constituent materials. 1) The first molecular layer is ferromagnetically coupled to the topmost FM layer and acquires a strong spin polarization of conduction states at $E_F$[7,9]. If the molecule carries a local magnetic moment (see the schematic of a phthalocyanine molecule and its central metal atom. Both are depicted by a red dot/arrow. Here, M=Mn with a magnetic moment of $3\mu_B$; Fe with $2\mu_B$, Co with $1\mu_B$ ; or Zn with $0\mu_B$.[19,24–26]), then this magnetic stabilization can be extended to subsequent molecular monolayers away from the interface[16] according to structurally imposed magnetic exchange interactions[19]. 2) The topmost FM layer's magnetism is altered relative to the FM's underlying layers. This magnetic hardening effect was first observed in magnetotransport experiments[13]. We disentangle how magnetic hardening and spinterface-stabilized molecular spin chains contribute to an effective exchange bias effect on the bulk portion of the FM thin film.

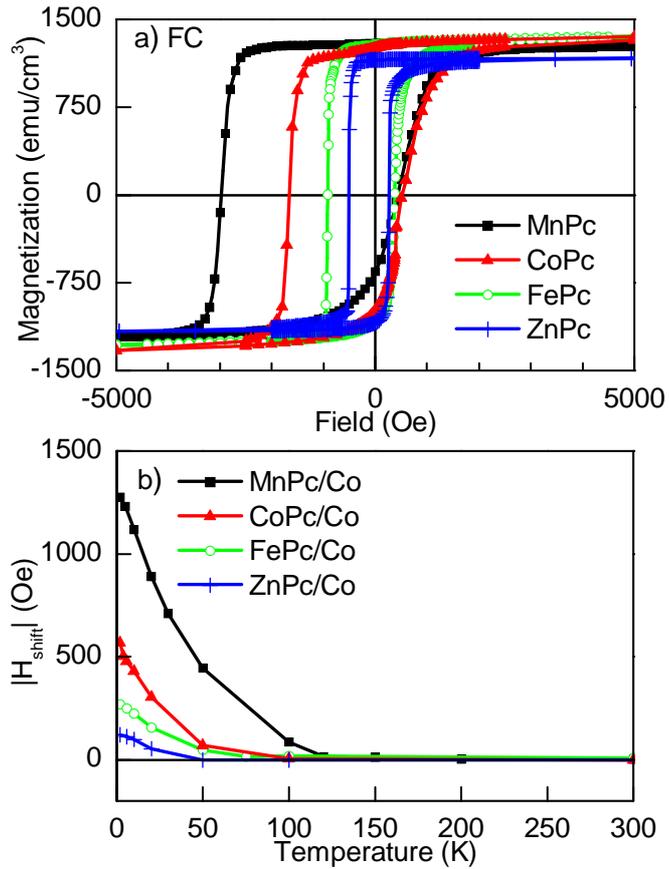

**Figure 2**. **Magnetic exchange effect of Co/MPc bilayers upon suppressing or enhancing the phthalocyanine molecule's magnetic moment.** Magnetization loops at 2 K of Au-capped Co(10 nm)/MPc(10 nm) bilayers after (a) field cooling (FC) at +3T. The shift in the loop center away from H=0 defines $H_{shift}$. (b) Temperature dependence of $H_{shift}$ for Co/MPc bilayers. A non-zero $H_{shift}$ is found for diamagnetic ZnPc, and increases as the M molecular site's moment is increased: it is one order of magnitude larger for MnPc.

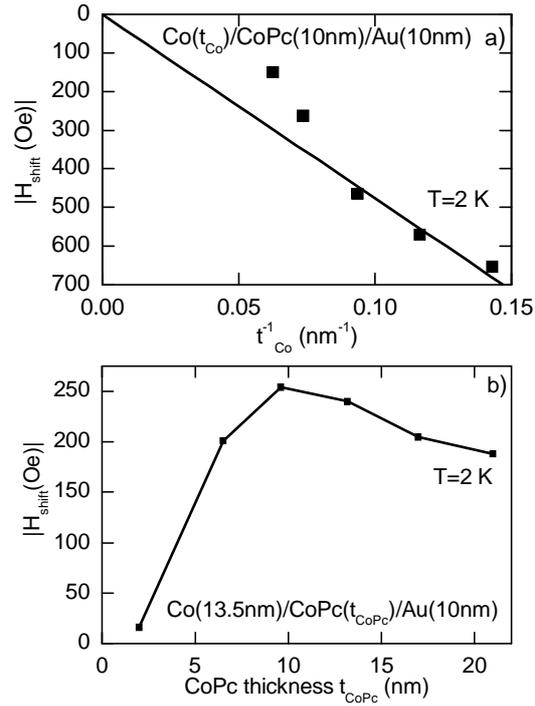

**Figure 3**. **Explicit proof that molecular spin chains contribute to the EB**. Dependence of $H_{shift}$ at T=2K upon varying (a) the Co thickness in Au-capped Co($t_{Co}$)/CoPc(10nm) bilayers and (b) the CoPc thickness in Au-capped Co(13.5 nm)/CoPc($t_{CoPc}$) bilayers. The $1/t_{Co}$ thickness dependence of $H_{shift}$ confirms the interfacial nature of the magnetic exchange effect from the standpoint of the FM layer's magnetism. The increase in $H_{shift}$ with increasing CoPc thickness confirms that molecular spin chains can enhance the effect generated by magnetic hardening[13] through an EB mechanism[16].

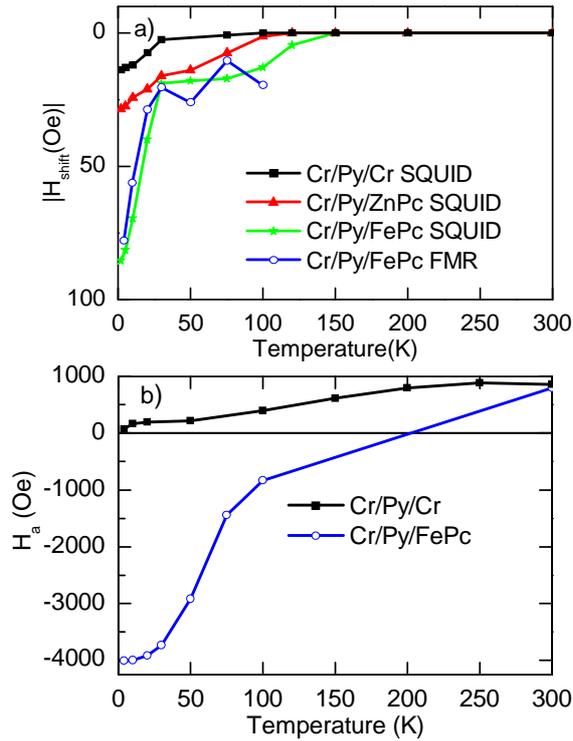

**Figure 4**. **Experimental confirmation of a uniaxial anisotropy within Py/FePc bilayers.** (a) Temperature dependence of $H_{shift}$ extracted from SQUID and FMR measurements for Au-capped Cr(15 nm)/Py(10 nm)/X(10 nm) with X = Cr, ZnPc, FePc. Similar data are found for FePc when using SQUID and FMR techniques. (b) Temperature dependence of the uniaxial magnetic anisotropy field $H_a$ deduced from FMR measurements for Cr/Py/Cr and Cr/Py/FePc films. A positive(negative) $H_a$ favors an out-of-plane(in-plane) easy magnetization axis. The substitution of FePc for Cr generates an additional, large negative contribution to $H_a$ whose temperature dependence mimicks that of $H_{shift}$.